\documentclass[aps,showpacs,preprintnumbers,amsmath,amssymb,
eqsecnum, twocolumn, tightenlines
]{revtex4}

\usepackage{graphicx}

\newcommand{\be}{\begin{eqnarray}}
\newcommand{\ee}{\end{eqnarray}}

 \newcommand{\gsim}{\mathrel{\hbox{\rlap{\lower.55ex \hbox {$\sim$}}
                   \kern-.3em \raise.4ex \hbox{$>$}}}}
\newcommand{\lsim}{\mathrel{\hbox{\rlap{\lower.55ex \hbox {$\sim$}}
                   \kern-.3em \raise.4ex \hbox{$<$}}}}

\def\N{${\cal N}\,\,$}

\newcommand{\ba}{\begin{eqnarray}}
\newcommand{\ea}{\end{eqnarray}}

\begin{document}


\title{ QCD-like theories with many fermions: \\ magnetic plasma and  unusual confinements }

\author { Edward Shuryak}
\address { Department of Physics and Astronomy, State University of New York,
Stony Brook, NY 11794}
\date{\today}

\begin{abstract}
This comment-style letter represents a part of my talks on
new developments in lattice QCD.  While it does not contain any new results, it 
containes some theoretical ideas and comparisons which, I think,
are not yet widely discussed in the lattice community. We point out that for $N_f\sim 10$ 
one can study a plasma phase  which is much more ``magnetic dominated" than
for small $N_f$. We also suggest
 certain tests/measurements
 to check if ``unusual confinement" phases  are or are not realized in this region.
\end{abstract}
\maketitle

\section{Magnetic plasma in many-flavor QCD?}

  The issue started with the  Dirac paper \cite{Dirac:1931kp} in which he famously observed that magnetically charged particles -- monopoles,dyons etc -- are possible,
provided the electric and magnetic coupling satisfy the Dirac condition 
\be \alpha_e \alpha_m= integer \ee
is satisfied, because only then the Dirac strings are invisible (pure gauge).  When Non-Abelian theories and color appeared, this condition gets modified but survived, with certain lattices of charges allowed
depending on the color group.

 Furthermore, it remarkably survived a generalization to quantum field theories and the running couplings. 
One extreme confirmation of this one may find e.g. in Seiberg-Witten treatment of the \N=2 super Yang-Mills theories
\cite{Seiberg:1994rs}. From their exact effective Lagrangian one can calculate the dependence
of the coupling on the location in moduli spaces, and see that at the points where monopoles/dyons
get massless the electric coupling tends to infinity, while the magnetic one goes to zero. 

Less extreme but still impressive  manifestation of the same idea
one finds in the finite temperature QCD. At high $T$ we are
at weak coupling, and thus it is made of electric particles (and thus called the  Quark-Gluon Plasma 
\cite{Shuryak:1978ij}).
As $T$ decreases and the electric coupling grows, the states which used to be solitonic and
exponentially suppressed start to become less and less suppressed. 

Finally, when effective electric and magnetic couplings are both the same $\alpha_e\approx \alpha_m\approx 1$,
 one expects the so called electric-magnetic equilibrium \cite{Liao:2006ry}: the electric and magnetic screening lengths and densities are about equal there. Using lattice data \cite{D'Alessandro:2007su} on monopole-monopole
 spatial correlation functions, Liao and myself
 \cite{Liao:2008jg}
 were able to extract the effective magnetic coupling $\alpha_m(T)$ as a function of temperature. We indeed find that
 it is the inverse of the electric one, so that monopole correlations get stronger as $T$ increases.
 The crossing of $\alpha_m(T)$ with $\alpha_e(T)$ had identified this equilibrium point: 
 in the pure gauge theory it happened to be
 around $T=1.4T_c$ . Thus we concluded that just above  deconfinement $T=T_c$
the plasma phase is a bit tilted toward the magnetic side already.


\begin{figure*}[t] 
   \centering
   \includegraphics[width=7cm]{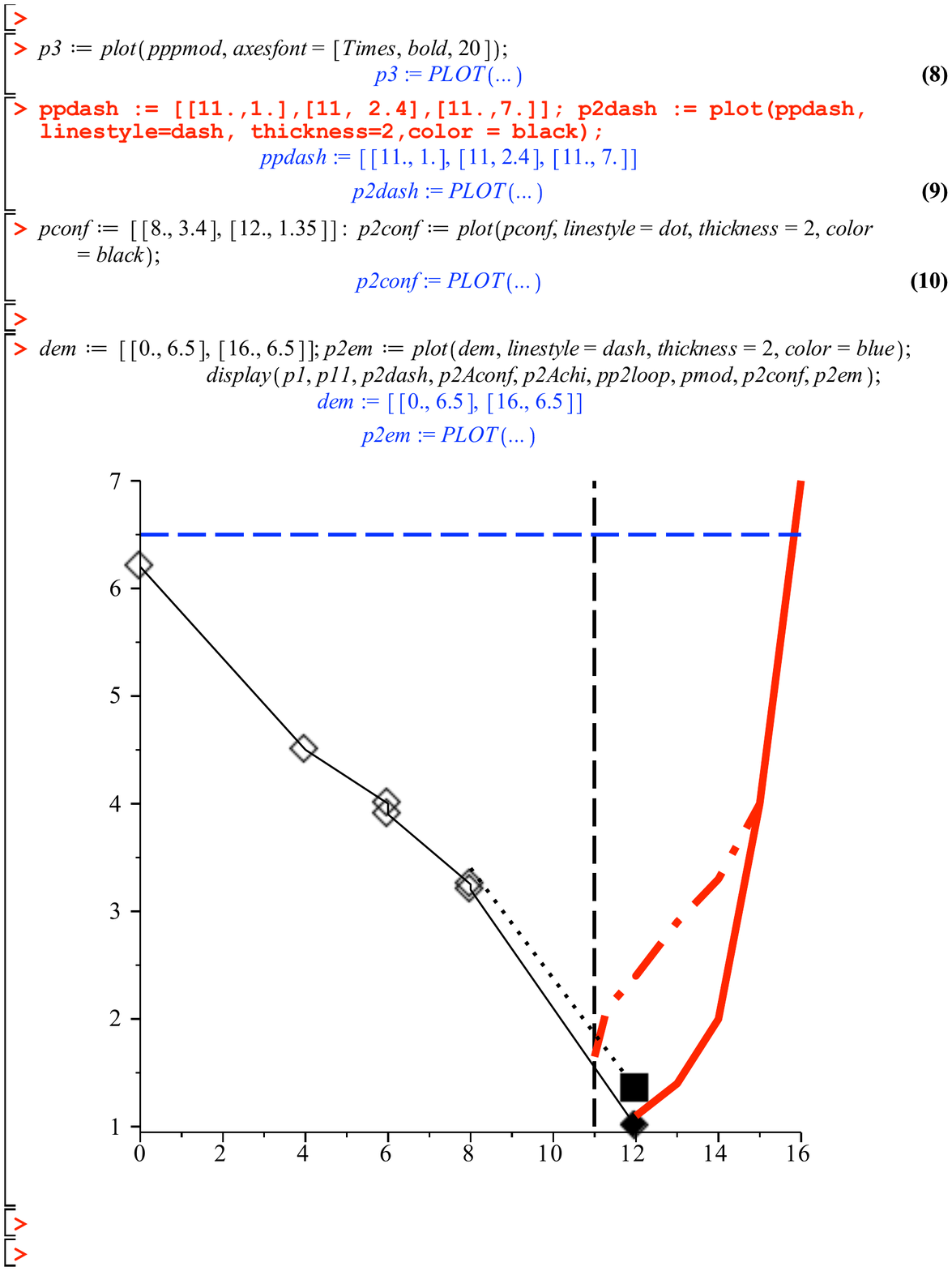}   
   \caption{ (Color online)   The critical lines for  chiral restoration (solid line and diamonds) and deconfinement (dotted line and a closed box point )  of the $N_c=3$ gauge theory.
 We plot the critical lattice coupling  at scale $T$ $\beta_c(T_c)=6/g^2_c(T_c)$ 
   versus  the number of fundamental quarks $N_f$.
     In   infrared fixed point  , calculated in the 2-loop approximation, is shown by the thick (red) lines. The vertical  dashed lines
separate ``conformal window": its location is a guess.  The (red)   dash-dotted  line is a guess for the actual location of the fixed point coupling.
The diamond points are from lattice studies \cite{Miura:2011mc}, and two points at 12 flavors are 
two phase transitions from \cite{Cheng:2011ic}. The picture itself is a slightly modified
phase diagram from \cite{Shuryak:2012aa}.   }
   \label{fig:romanplot}
\end{figure*} 

The reason we repeated  those well established results is related with recent progress of
supercomputer technology, which made accessible to lattice QCD practitioners new region of the
phase diagram, with $N_f=4..16$ (for $N_c=3$ most actively studied).  

In Fig.\ref{fig:romanplot} we show the phase diagram, using as variables an effective coupling at scale $T$ vs the number of fermion flavors $N_f$.
High $T$ is weak coupling, thus it corresponds to the top of the digram. Going downward, at fixed $N_f$,
corresponds to decrease of the temperature. Blue horizontal dashed line indicates the approximate position
of the E/M equilibrium on this plot. (The bare lattice coupling and effective true one are not the same!)

At the left side of the plot, if one is going downward to stronger coupling, one eventually meets 
thin solid line (connecting the diamond points), which  corresponds to the chiral
symmetry breaking transition. Below this line one also observes confinement:
so it is the usual hadronic phase, the same as we live in in the real world QCD.

The right side of this plot has (red) thick line which asymptotes vertically at $N_f=11N_c/2=16.5$,
the value at which the asymptotic freedom is lost. This  line is the critical coupling calculated in 2 loops:
 the dash-dotted line is my schematic guess of how the exact critical IR line goes. In the part of the diagram
 to the right of the vertical dashed line the cooling of the plasma $T\rightarrow 0$ ends in the IR at this line, so that the region below it is unreachable from above.
 This part of the diagram is called {\em the IR-conformal} region, and it is interesting to study in its own right.
 
 One obvious observation is that 
 in the region $N_f\approx 11$ the corresponding phase transitions shifts to extremely low temperatures, or,
in other words, to much stronger coupling, as compared to the ``real world QCD" with smaller $N_f=2..3$.
The horizontal dashed line at the top of the figure roughly indicates the electric-magnetic equilibrium line,
according to \cite{Liao:2008jg}. The effective coupling for the coolest plasma on the plot, at $N_f\approx 11$
is about factor 3 higher than for $N_f=0$ theory.
Thus  the effective electric and magnetic couplings seen in interactions are
$\alpha_e\sim 3$ while the magnetic one $\alpha_m\sim 1/3$. This is ``the most magnetic" plasma
one can manifecture (on supercomputers). 

  There are many obvious questions one can ask about it.
Is it as good a liquid, as seen in the real-world QGP at RHIC and LHC, at $T=(1-2) \,T_c$? Perhaps not,
as monopoles are in it quite weakly coupled.  What are its effective degrees of freedom?
That should
depend on spectroscopy of magnetic objects we don't understand (see also below).
   What is their interactions and an effective QFT describing both electric and magnetic 
   objects can or cannot be formulated? 

(To the reader who noticed the dotted line and other kinds of data points: we return to them later.)

\section{Confinement and unusual confinements}
  Confinement is traditionally viewed as a ``dual superconductivity" phenomenon \cite{Mandelstam:1974pi,'tHooft:1981ht}, induced by
   Bose-Einstein Condensation (BEC)
of certain  magnetically charged bosons. In theories with adjoint scalar fields, such as the \cal{N}=2 SYM \cite{Seiberg:1994rs}, 
these objects has been
identified with 't Hooft-Polyakov monopoles and the confinement mechanism is under rather firm theoretical control. 

In QCD-like theories the situation is different: the debates about the nature and quantum numbers of such monopoles continue to this day. 
The problem is there is no obvious adjoint scalar which can provide
appropriate ``Higgsing" .
At $T>T_c$ the so called Polyakov line \be P= {1\over N_c} Tr \, exp(i \int A_4 dx_4) \ee has a nonzero vacuum expectation value: while 
not being a scalar, it has an adjoint color and provides a ``Polyakov gauge" in which the diagonal
Abelian subgroup remain massless. 
Lattice monopole has been identified in this as well as Maximal Abelian gauges. The number of those which
makes at least one period over the time variable seem to be lattice-independent and 
show rapid growth of the monopole density as $T_c$ is approached from above \cite{D'Alessandro:2007su}. 
Presence of the monopole condensate at $T<T_c$ has been shown by the Pisa group \cite{Cossu:2008wh}.

More recently, it has been demonstrated in \cite{D'Alessandro:2010xg}
that such lattice monopoles do show bose-clustering very similar to that of any system undergoing BEC,
with divergence of the cluster size falling (inside small error bars) at $T_c$. 
Thus, the BEC nature of the ``usual" confinement seem to be confirmed, in spite of the fact that
   the exact nature of the magnetically charged objects remains to be studied further.

 In theories with  massless fermions the monopoles have normalizable fermionic zero modes, interpreted as 
the energy levels which may or may not be occupied. This leads to a zoo of $\sim 2^{2N_f}$ magnetic
charge-1 objects 
(extra 2 in exponent is due to antiquark modes, which exists as well for opposite chirality).
For  QCD-like theories with $N_f\sim 10$ under consideration now, the number of those states is truly huge.
Furthermore, there are interesting two-monopole bound states (see below). Ultimately, all of those
have to be included in the partition function of the ``magnetic plasma", with many fermions and strong coupling.

 The spectroscopy of those states in the non-supersymmetric theories is not yet studied: but it can be, 
at least in a weak coupling limit using semiclassical approach. If one however wants 
 to understand what happens near $T_c$, when the coupling gets strong
and the objects in question becomes near-massless or even tachionic, lattice remains the only approach
available.
 
 Let me now ask a general question  whether  (i) {\em  only one type} of the resulting magnetic objects
 is Bose-condensing on the whole phase diagram (presumably  the minimal ``fermion-empty" monopole, with  $n_m=1$); 
 or  (ii) under different conditions (that is points on the phase diagram or $T,N_f$)
  various states with {\em different quantum numbers}
 may undergo BEC as well. The latter case I call the {\em ``unusual confinements"}.
  In the latter case, one would expect to find the phase transitions between   
different confining phases, if the order parameters of different confinement phases are not the same.

  Spectroscopy of magnetic states is under better control in 
supersymmetric theories. The maximally symmetric $\cal{N}$=4 theory 
has magnetic supermultiplet of 16 states, with the same spins and quantum numbers as electric theory.
(Thus running of $g$ and $1/g$ is give by the same Lagrangian, with the only conclusion of no running at all.

As shown by Sen \cite{Sen:1994yi}, in this case $LS(2,Z)$ symmetry require existence of
the  bound states with any charges $n_m,n_e$ provided they are mutually prime. He also explicitly
found such state for $n_m=2,n_e=1$ using   Atiyah-Hitchin two-monopole moduli space metric.

Due to celebrated Seiberg-Witten works $\cal{N}$=2 QCD is  also under control. I especially point out
the supersymmetric QCD with fundamental quarks/scalars  
 \cite{Seiberg:1994aj} near its conformal window.  Internal consistency of the whole scenario, as fermions one by one change their mass
 from large to small, had allowed to fix uniquely the quantum numbers of $all$  magnetic objects
 which get massless on the moduli spaces (and produce BEC if supersymmetry is broken).
  Because of additional
 gluino species and their contribution to the beta function, addition of
massless quarks are restricted to $N_f=1,2,3$, for the number of colors $N_c=2$
discussed in their work. (Four flavors is already fully conformal theory.)

Let me mention only one case, with the largest $N_f=3$ quark number which has asymptotic freedom. 
Seiberg and Witten predicted two district singularities
on the moduli space, which correspond to the following particles becoming massless: \\
(i)  a quartet  of states with magnetic charge $n_m=1$ and electric charge $n_e=0$;  and (ii) a singlet with $n_m=2,n_e=1$.   Various SUSY breaking terms would transform those singularities into two non-equivalent vacua,
with two confinement
phases: \\ (i) the former $with$ the chiral symmetry breaking, and  (ii) the latter one $without$ it. 

 Both are examples of what I call the ``unusual confinements", as the objects undergoing BEC 
 have quantum numbers different from being just ``fermion-empty" monopoles of charge 1.
 The former is a monopole on which certain quarks ``get a ride" .
 (Recall that adding together the spin 1/2 of a quark with its spin-1/2 color in SU(2)
  one finds zero modes with grandspin=0.
 This makes them bosons, ready to BEC.) 
 
 The second one is even more interesting: it is also a ``molecular" state made of two monopoles
 bound together, line Sen's state, but now by 3 quarks. It has been shown in \cite{Gauntlett:1995fu} 
 using appropriate index theorem and properties of the Atiyah-Hitchin geometry of the
 two-monopole space of relative motion, that only starting with $N_f=3$ there is one (and only one)
 such bound state.
 This must be the state $n_m=2,n_e=1$ with the singlet flavor,  predicted by Seiberg and Witten.
 As far as I was able to trace it, nobody actually found its wave function in this theory.
 (Of course, supersymmetry requires the fermionic excitation energy to be non-negative,
 and index theorems only count the zero-energy ones.)

   Let us now return to the phase diagram in Fig.1. Recently it has been shown  in \cite{Cheng:2011ic} (see also \cite{Miura:2011mc,Deuzeman:2010gb}) that at 8 and 12 flavors there appear two separate phase transitions,
with a new phase in between the solid and dotted lines. It is confining
 while the chiral symmetry remains unbroken. 
(The two points at 12 flavors and the dotted deconfinement line corresponds to these results.
 I don't have the data points to plot yet for 8 flavors.)

Is it possible that  there is a  relation between one (or even both)
 phases observed in those lattice studies with  the unusual confinements in the
 $\cal{N}$=2, $N_f=3,N_c=2$ supersymmetric QCD?

Theoretically, the underlying dynamics  remains  very unclear: various interactions may
   pick up a certain states to BEC out of large zoo
of states available in many different ways. Fortunately, the supercomputers had already solved
the problem in question: all one needs to do is to analyze the output. 
The testing includes the identification of the magnetic objects which undergo BEC in
both phases and {\em check if they have any unusual/additional quantum numbers}.

Another indication to an unusual phases would, of course, be some phase transition lines
separating them and the ``usual" confinement at smaller $N_f$.
This would however  require new simulations with variable masses to fill the phase diagram continuously,
which is far from being the case at the moment.

Even if those tests would turn out to be negative, finding that the confining
phases at large $N_f$ are only the same ``usual" monopoles, one should still be able
to find all  magnetic objects  present in the thermal ensemble right above $T_c$,
provided they are light enough to be present in sufficient quantities. 
We hope to provide some estimates for that elsewhere \cite{Liao_now}
 
\vskip .25cm {\bf Acknowledgments.} 
I  thank my former collaborators on related papers, Massimo D'Elia and Jinfeng Liao,
with whom I discussed these issues many times. I learned a lot from
discussions with Nick Dorey,Anna Hasenfratz and Edward Witten. This material was in my
talks at Bari conference in September 2011, at
 KITP, Santa Barbara and IAS,Princeton in spring of 2012: I acknowledge
support I obtained from all these institutions.


\end{document}